\documentclass{elsart}
\usepackage[authoryear]{natbib}
\usepackage{graphicx}
\usepackage{amssymb}
\usepackage{amsmath}
\usepackage{url}

\begin{document}
\begin{frontmatter}
\title{Coexistence and invasibility in a two-species competition model
with habitat-preference} \author[NBI,CO]{Simone Pigolotti} \and
\author[ISC,RM]{Massimo Cencini} \address[NBI]{The Niels Bohr
International Academy, The Niels Bohr Institut, Blegdamsvej 17,
DK-2100 Copenhagen, Denmark} \address[CO]{Corresponding author, email:
pigo@nbi.dk, Ph. +4535325238, Fax +4535325425} \address[ISC]{Istituto
dei Sistemi Complessi (ISC), CNR, Via dei Taurini, 19 00185 Rome,
Italy} \address[RM]{Dipartimento di Fisica, University of Rome
``Sapienza'', Piazzale A. Moro 5, 00185 Rome, Italy}
\begin{abstract}
The outcome of competition among species is influenced by the spatial
distribution of species and effects such as demographic stochasticity,
immigration fluxes, and the existence of preferred habitats. We
introduce an individual-based model describing the competition of two
species and incorporating all the above ingredients. We find that the
presence of habitat preference --- generating spatial niches ---
strongly stabilizes the coexistence of the two species. Eliminating
habitat preference --- neutral dynamics --- the model generates
patterns, such as distribution of population sizes, practically
identical to those obtained in the presence of habitat preference,
provided an higher immigration rate is considered.  Notwithstanding
the similarity in the population distribution, we show that
invasibility properties depend on habitat preference in a non-trivial
way. In particular, the neutral model results results more invasible
or less invasible depending on whether the comparison is made at equal
immigration rate or at equal distribution of population size,
respectively.  We discuss the relevance of these results for the
interpretation of invasibility experiments and the species occupancy
of preferred habitats.
\end{abstract}


\end{frontmatter}

\section{Introduction}

A central problem in community ecology is to understand the ecological
forces leading to the observed patterns of coexistence or exclusion of
competing species \citep{Ricklefs1993,Brown1995}. This issue is
important for understanding both simple communities made up of few
species \citep{Chesson2000} and ``biodiversity hotspots'' with a large
number of coexisting species \citep{Leigh2004}.  Historically, this
problem has been approached at two distinct levels. On the one hand,
focus has been put on the detailed mechanisms of interaction between
species (e.g. intra- and inter-specific competitions) caused by their
differentiation in exploiting resources, resulting in the concept of
the ecological niche \citep{Chase2003}. For example, it has been shown
how habitat heterogeneity \citep{Beckage2003} or a tradeoff in
dispersal range strategies \citep{Bolker1999} may promote
coexistence. An alternative explanation for the observed species
richness and distribution is in terms of processes intrinsically due
to chance, such as colonization, immigration and extinction
\citep{Macarthur1967}, disregarding differences among species.

In recent years, the neutral theory of biodiversity
\citep{Hubbell1979,Bell2001,Hubbell2001} considerably developed the
latter approach by explicitly assuming equivalence among species at
the individual level. The interest in the neutral theory has been
triggered by its ability to successfully predict several biodiversity
patterns observed in tropical forests, such as species-abundance
distributions in different permanent sampling plots
\citep{Bell2001,Hubbell2001,Volkov2003} and species-area relations
\citep{Durrett1996,Bell2001,Hubbell2001}. Its success underlined the
importance of stochasticity (ecological drift) and dispersal
limitation in the assemblage of natural communities
\citep{Chave2004,Alonso2006}, which are now recognized as key
ingredient also in niche-based models \citep{Tilman2004}.  However,
niche-based models yield predictions for the biodiversity patterns
which perform similarly to neutral ones when compared with data
\citep{Chave2002,McGill2003,Mouquet2003,Gilbert2004,Tilman2004}. This
suggests that these patterns tend to average out the dependence on the
details of the theory (see also the discussion in \citet{Pueyo2007})
and thus cannot be used to discriminate the relative importance of
niche-based and neutral forces. In this perspective, the study of
dynamical properties such as invasibility can be a promising way to
disentangle these effects \citep{Daleo2009}.

To understand the key differences between neutral and non-neutral
competition, it is useful to consider models that can be continuously
tuned from niche-based to neutral settings by varying some parameters
\citep{Chave2002,Gravel2006, Adler2007}.  An obvious difficulty with
this program comes from the unavoidable complexity of realistic
niche-based models \citep{Chase2005}, where species are not equivalent
and the environment is heterogeneous both in space and time.  This
suggests an approach whereby simplified models with few parameters are
studied, for example by making some specific assumptions on how
neutrality is violated.

In this paper we study the dynamics of two species $A$ and $B$ that
compete for space. The model is devised in such a way that a single
parameter controls the overlap between the niches occupied by the two
species, from complete -- neutral -- to no overlap -- two independent
niches. The model is individual-based and incorporates the basic
ingredients of neutral theory: coexistence results from immigration
from a metacommunity, balancing demographic stochasticity, which alone
would lead to extinction.  Niches are introduced in this neutral
scenario via preferential habitats: half of the sites in the ecosystem
are favorable for the colonization of individuals of one species and
the other half are favorable for the other species.  The ecological
advantage is realized through a biased ``lottery''
\citep{Chesson1981}. We consider a symmetric situation by choosing the
same statistical bias, $\gamma$, for individuals of species $A$ and
$B$ to colonize their respective preferred habitats. When $\gamma=0$
(no habitat diversification) the model reduces to the voter model
\citep{Holley1975,Cox1986}, which is a prototype of neutral
dynamics. Increasing $\gamma$, species acquire an advantage in
colonizing some sites and a disadvantage in others. A very large
$\gamma$ eventually leads to segregation of the two species to their
preferential habitats.  Segregation will be complete when the choice of
dispersal allows individuals to reach all their preference sites or
incomplete in the presence of dispersal limitation.

The aim of this work is to use this simple model to understand the
effect of habitat diversification on coexistence and dynamics of
ecological communities. In particular, our concern will be on
contrasting the effect of habitat diversification with the neutral
model where no preferred habitat exists.

\section{Model}

We consider an individual based, spatially-explicit model of a
community made of two competing species $A$ and $B$ with population
$N_A$ and $N_B$, respectively. The community lives in a patch made of
$N=L^2$ sites on a square lattice of side $L$, on which we assume
periodic boundary conditions. Each lattice site is occupied by a
single individual of one of the two species. For the sake of
simplicity, we assume that the patch is saturated, i.e. with no empty
sites --- each dead individual is immediately replaced, so that the
total number of individual is constant, $N_A+N_B=N$. The latter
hypothesis is commonly assumed for its convenience
\citep{Hubbell2001,Chave2002} and, strictly speaking, corresponds to
considering infinite fecundity. However, a finite but reasonably high
fecundity would lead to almost-saturated ecosystems with qualitatively
similar dynamics \citep{Durrett1996,Chave2002}.

\begin{figure}[t!]
\centering
\includegraphics[width=.4\textwidth]{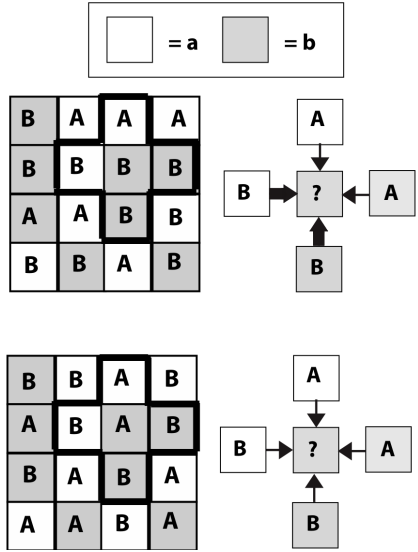}
\caption{Sketch illustrating the model with (top) and without (bottom)
  habitat preference.  Two representative configurations of $4\times4$
  systems are shown, white squares are advantageous to $A$, gray ones
  to $B$, and on the right we sketch the lottery dynamics
  (eqs. (\ref{ratesa} and (\ref{ratesb})): the width of arrows
  represents the habitat preference intensity $\gamma$. Notice that in
  the bottom panel the width of arrows is insensitive to the site
  specificity\label{fig1}}
\end{figure}

Our main interest here is to study the effect of habitat preference on
competition.  To this aim, we assign to each site a specificity: half
of the sites are favorable (as below specified) to individuals of
species $A$ and the other half to individuals of species $B$. We
denote such sites by $a$ and $b$, respectively.  The site specificity
can have several different (often concomitant) ecological origins such
as abundance of resources, predation pressure \citep[see, e.g., the
review by][]{Amarasekare2003}, and/or environmental conditions such as
elevation, temperature, soil moisture or other parameters as in
\citet{Zillio2007} and as suggested by observations
\citep{Beckage2003}. The net effect of these different mechanisms is
here assumed to increase by a factor $\gamma$ the chance of
individuals to colonize a preferred site. This is illustrated in the
top panel of Fig.\ref{fig1}, to be compared with the bottom cartoon
which shows the neutral model, without site specificity.  Similar
models have been proposed also in the context of heterogeneous
catalysis \citep{Redner1} and social dynamics \citep{Redner2}.

For the sake of simplicity, site specificity is randomly assigned at
the beginning and left unchanged during the dynamics.  Of course, in
natural ecosystems, spatial arrangement of sites with a certain
specificity will usually be characterized by a certain degree of
correlations, which will in general tend to enhance niche effects.  In
this respect, we expect that our choice will tend to underestimate the
effect of habitat preference. Clearly, the model can be generalized by
introducing asymmetries, i.e. different $\gamma$'s for the two species
or different fractions of advantageous sites.  Here we shall limit our
analysis to the simple symmetric case, so that no species has a net
advantage and the degree of habitat preference is controlled by a
unique parameter.

We also assume a continuous immigration in the patch of individuals
$A$ or $B$ at rate $\nu$.  This inflow is necessary to avoid the
drift to extinction of one of the two species.

For any given size $L$ of the patch, which fixes the number of
individuals $N=L^2$, the model is controlled by two parameters only:
the colonization advantage $\gamma$ and the immigration rate
$\nu$. The elementary step of the dynamics is as follows
\begin{description}
\item{i)} a site is randomly chosen and the individual there residing is
  killed;
\item{ii)} with probability $(1-\nu)$, the individual is replaced by a
  copy of one of the four neighbors, chosen via a lottery which gives
  a competitive advantage (modeled as a weight $\gamma$) to
  individuals having that site as preferred habitat (see the sketch in
  Fig.\ref{fig1} and eqs. (\ref{ratesa}) and (\ref{ratesb}) below).
\item{iii)} with probability $\nu$, the individual is replaced by an
  immigrant. For simplicity, we assume the two species being
  equipopulated at the metacommunity level, so that the probability of
  being replaced by an individual of one of the two species is $1/2$,
  apart from the competitive advantage on the specific empty site.
\end{description}

In formulas, steps \textit{ii)} and \textit{iii)} can be expressed as
follows.  If the individuals is killed in a site of type $a$
advantageous for individuals of species $A$, the probabilities of
being replaced by an individual $A$ or $B$ are given by
\begin{equation}
\label{ratesa}
\begin{array}{ll}
  W^a_A(n_A,n_B)=&(1-\nu)\frac{(1+\gamma)n_A}
{(1+\gamma)n_A+n_B}+\nu\frac{1+\gamma}{2+\gamma}\\
&\\
  W^a_B(n_A,n_B)=&(1-\nu)\frac{n_B}{(1+\gamma)n_A+n_B}+
\nu\frac{1}{2+\gamma}\,,
\end{array}
\end{equation}
respectively, where $n_A$ and $n_B$ denote the number of individuals of species
$A$ and $B$ in the neighborhood of the considered site. Similarly, if the
individual is killed in a site of type $b$, we have
\begin{equation}
\label{ratesb}
\begin{array}{ll}
  W^b_A(n_A,n_B)=&(1-\nu)\frac{n_A}{n_A+(1+\gamma)n_B}+ 
\nu\frac{1}{2+\gamma}\nonumber\\
&\\
  W^b_B(n_A,n_B)=&(1-\nu)\frac{(1+\gamma)n_B}{n_A+(1+\gamma)n_B}
+\nu\frac{1+\gamma}{2+\gamma}\,.
\end{array}
\end{equation}
The competitive lottery (\ref{ratesa}) and (\ref{ratesb}) used in step
\textit{ii)} represents a biased (non-neutral) generalization of that
used in neutral models \citep{Hubbell2001}. Indeed, for $\gamma=0$,
the neutral dynamics of the voter model with only two species is
recovered \citep{Holley1975}. 

Notice that the model is set up in such a way that the fitness
advantage to be on colonization: after having colonized a site,
mortality and dispersal do not depend on being on a preference
site. In other terms, the fitness advantage belongs to the seeds and
not to the individuals themselves. In this interpretation, and at
variance with \citet{Chesson1981}, we excluded from the contribution
to the (implicit) seed pool of the individual who died. Although the
latter may be more realistic in modeling, e.g., perennial plants
\citep{Lin2009}, we preferred the first to directly compare the
neutral version of the model ($\gamma=0$) originally proposed by
\citet{Hubbell1979} with the non-neutral ($\gamma>0$) variants. We do
not expect big differences in the outcome of the two models, a part
from a weak discrepancy in the percentage of occupancy of preferential
habitats. We also note that this difference becomes less relevant when
longer dispersal is introduced.  Indeed, the model can be generalized
to different form of dispersal mechanisms, e.g. with a given finite
range. Previous investigations on similar models
\citep{Rosindell2007,Zillio2007,Pigolotti2009} have shown that
different dispersal mechanisms yield qualitatively similar results, as
far as they are finite-ranged and all species adopt the same dispersal
strategy. Here, we consider the two limiting situations of
nearest-neighbor and global dispersal.

In the global dispersal case, all individuals present in the patch can colonize
the site left empty by the dead one. This limiting case strongly
simplifies the simulations of the model since it makes the distance
and spatial distribution of the sites irrelevant. The global dispersal
model may be thought as a two islands model \citep{Wright1931}, where
each of the two islands has room for $N/2$ individuals and contains
all the sites favorable to one of the two species. The state of the
system is then unequivocally identified by the numbers $N_{Aa}$ and
$N_{Bb}$ of individuals of the two species occupying the respective
sites of preference.  All the other variables can be expressed in
terms of $N_{Aa}$ and $N_{Bb}$, for instance: the number of
individuals of species $A$ (resp. $B$) outside their preference sites
is $N_{Ab}=N/2-N_{Bb}$ (resp. $N_{Ba}=N/2-N_{Aa}$) and the total
number of individuals of species $A$ (resp. $B$) is
$N_A=N/2+N_{Aa}-N_{Bb}$ (resp. $N_A=N/2+N_{Bb}-N_{Aa}$). The evolution
of the system is thus determined by the probabilities per elementary
step that $N_{Aa}$ and $N_{Bb}$ increase or decrease by one unity:
\begin{eqnarray}
\label{eq:globaldisp}
  \mathcal{W}_{N_{Aa}\rightarrow N_{Aa}+1}&=& \left(\frac{1}{2}-\frac{N_{Aa}}{N}
\right) \ W_A^a(N_A,N_B) \nonumber\\
    \mathcal{W}_{N_{Bb}\rightarrow N_{Bb}+1}&=& \left(\frac{1}{2}-\frac{N_{Bb}}{N}
\right) \ W_B^b(N_A,N_B)\nonumber\\
    \mathcal{W}_{N_{Aa}\rightarrow N_{Aa}-1}&=& \frac{N_{Aa}}{N}\ 
W_B^a(N_A,N_B) \nonumber\\
    \mathcal{W}_{N_{Bb}\rightarrow N_{Bb}-1}&=& \frac{N_{Bb}}{N}\  W_A^b(N_A,N_B).
\end{eqnarray}
where $W^x_Y$ is given by eqs. (\ref{ratesa}) and (\ref{ratesb}) with
$n_A$ and $n_B$ replaced by $N_A$ and $N_B$, respectively.

For $\gamma=0$, the model is neutral and reduces to the ordinary voter
model with only two species and immigration (the multi-species version
was considered in \citet{Durrett1996}) in the spatially explicit case,
or to the Moran model with mutation \citep{Moran1958} in the global
dispersal version. Conversely, when $\gamma$ becomes very large, the
competitive advantage gets so intense that individuals tend to
localize in their preferred habitats, with an extremely low chance of
colonizing the rest of the ecosystem, so that the two species do not
compete anymore.

\section{Results}

\bigskip
\centerline{\textit{Extinction times and fixation probabilities}} 
\medskip
In the absence of immigration ($\nu=0$) and for any choice of $N$,
$\gamma$ and dispersal range, persistent coexistence is not possible,
since species cannot recover from a local extinction event.
Ecological drift will ultimately drive one of the two species to
extinction, and the dominance of the other will become stable --- reaching
\textit{fixation} as from population genetics terminology, see
e.g. \citet{Gillespie1994}. Studying the properties of the dynamics
toward extinction is however interesting and informative
\citep{Chesson1981,Chesson1982}. In particular, the time needed for
the extinction of one of the species is important to understand how
crucially biodiversity depends on a steady immigration flux.
Moreover, the way the extinction time and the probability of fixation
of one of the two species depend on the deviation from neutrality
allow to quantify to what extent habitat preference promotes
coexistence.

\begin{figure}[t!]
\centering
\includegraphics[width=0.45\textwidth]{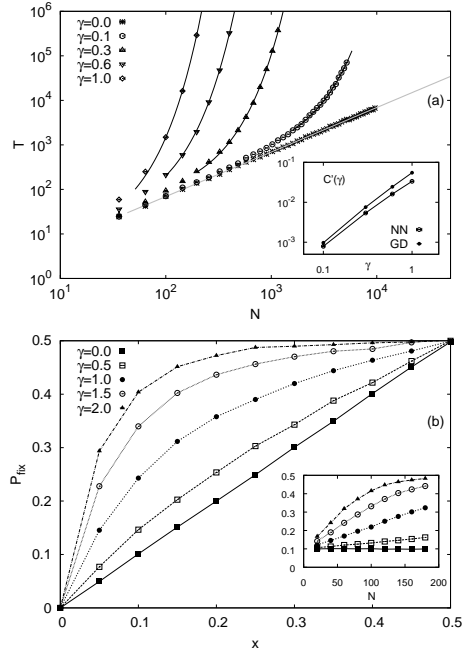}
\caption{Results for the model with global dispersal without
immigration ($\nu=0$): (a) Mean extinction times $T$ as a function of
$N$ for different $\gamma$, as in label. The black lines indicate
exponential fits of the form $T= C \exp(C'(\gamma) N )$ while the gray
straight line indicate the neutral expectation $T \propto N$. The
inset shows $C'(\gamma)$ versus $\gamma$ for both the global dispersal
(GD) and nearest neighbor (NN) models. (b) Probability of fixation
$P_{fix}$ of $A$ vs the initial fraction $x=N_A/N$ for $N=40$. Inset:
$P_{fix}$ vs $N$ holding fixed $x=N_A/N=0.1$.}\label{figtempi}
\end{figure}

We start describing a set of simulations in which the two species are
equipopulated at the initial time, i.e. $N_A=N_B=N/2$, and individuals
are randomly placed in the system. The dynamics is then followed until
the extinction of one of the species. By averaging over many
realizations of the dynamics (from $5\times10^3$ to $10^4$), we can
access the extinction time. Here we focus on the average time, $T$,
though also the fluctuations have an ecological relevance
\citep{Pigolotti2005}.  The average extinction time $T$ as a function
of the community size $N$ and for several values of $\gamma$ is shown
in Fig.\ref{figtempi} as obtained with the global dispersal model. As
customary in models with overlapping generations, we measure $T$ in
generations, where the time unit corresponds to $N$ successive time steps
\textit{i)-iii)} described in the {\em model} section.  For
$\gamma=0$, we recover the result expected for the Moran model
\citep{Moran1958} i.e. $T\propto N$. While the presence of habitat
preference $\gamma>0$ leads to a dramatic increase of the average
extinction time: for large enough populations, $T$ becomes
exponentially large in the community size $N$, i.e.
$$
T\sim C \exp(C'(\gamma) N )
$$ where $C'$ depends on $\gamma$ (as shown in the inset of
Fig.\ref{figtempi}a). The exponential dependence implies that the
coexistence may be considered stable for large $N$
\citep{Chesson1982}.  The spatially explicit version of the model (not
shown here) presents the same qualitative features with small
quantitative changes: in the neutral case $\gamma=0$, logarithmic
corrections are present $T\propto N\log(N)$ \citep{Krapivsky1992} and
the exponential rates for $\gamma>0$ are slightly different from those
obtained with global dispersal (inset of Fig.\ref{figtempi}a). In
large communities, even a tiny habitat preference leads to a dramatic
increase in the average extinction time, stabilizing the system on any
realistic timescale.

To further confirm the stabilizing effect of habitat preference, we
consider a different setting in which species $A$ is present at
initial time with a fraction of individuals $x=N_A/N$. By averaging
over $10^5$ realizations for each $x$, we computed the probability
that species $A$ becomes fixated, $P_{fix}(x)$ (Fig.\ref{figtempi}b).
In the neutral case, $P_{fix}(x)=x$ \citep{Gillespie1994}, while
$P_{fix}(x)$ is closer to $1/2$ when $\gamma$ is increased, the effect
being stronger the larger is the community size (see inset).
Therefore, strong habitat preference tends to compensate any initial
disproportion between two large populations, making equally likely the
fixation of one of the two species.  In particular, if $A$ is the
invading species, i.e. the less represented at the beginning ($x\ll
1/2$), its chances of invading the systems greatly increase in the
presence of habitat preference, in agreement with other models
and observations \citep{Melbourne2007}.

\begin{figure}[b!]
\centering
\includegraphics[width=.45\textwidth]{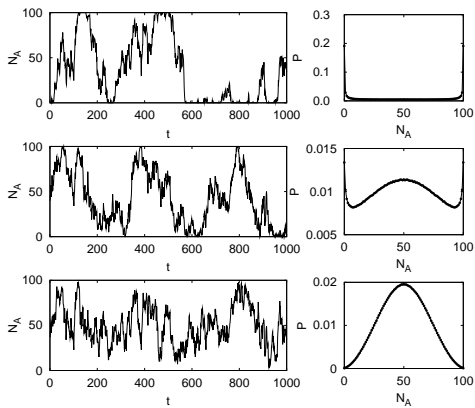}
\caption{Different regimes of coexistence in the presence of
  immigration in the global dispersal model with habitat preference:
  evolution of the population $N_A$ across $10^3$ generation for a
  system of $N=100$ individuals (left) and the corresponding
  distributions $P(N_A)$ (right). The three panels are
  obtained holding the habitat-preference intensity fixed $\gamma=0.3$ and
  varying the immigration rate $\nu$: (top) $\nu=0.003$ corresponding
  to monodominance; $\nu=0.015$ leading to a mixed phase; $\nu=0.05$
  displaying coexistence (see text for details).}
\label{figtsdistr}
\end{figure}

\bigskip
\centerline{\textit{Regimes of coexistence}} 
\medskip

The presence of an immigration pressure allows species to recover from
local extinction events, ensuring dynamical coexistence, as
illustrated in the left panels of Fig.~\ref{figtsdistr}, where the
evolution of the population $N_A$ is shown over $10^3$ generations at
varying the rate of immigration $\nu$. The figures refer to the global
dispersal model with a habitat-preference intensity $\gamma=0.3$. We mention
that qualitatively similar features are observed also in the
nearest-neighbor dispersal case and also when the model is neutral.

As expected from classical theories \citep{Macarthur1967}, increasing
the immigration pressure enhances the degree of coexistence of the two
species. At fixed $N$, if $\nu$ is small, one of the two species
dominates for most of the time.  Stochastic fluctuations, whose
amplitude decreases at increasing $N$, lead to an alternation in the
dominating species (top panel of Fig.~\ref{figtsdistr}).  The turnover
in the dominating species takes place on timescales rapidly growing
with $N$ (roughly of the order of the average extinction time). At
intermediate values of $\nu$, coexistence is possible, but episodic
local extinctions can rule out a species from the system for several
generations (middle panel). Finally, increasing $\nu$ even further,
local extinctions cannot persist and coexistence becomes the rule: it
is more and more likely to observe states in which the two species are
roughly equipopulated (bottom panel).  We quantify these three
behaviors in terms of the functional shape of the probability
$P(N_A)$, averaged over time, of one of the two species to occupy a
given fraction of the ecosystem \citep[see, e.g.][]{Loreau1999}. We
recall that there is a complete symmetry between $A$ and $B$ so that
$P(N_A)=P(N-N_A)=P(N_B)$.  In particular, with reference to the right
panels of Fig.~\ref{figtsdistr}, from top to bottom we can distinguish
the following classes of distributions corresponding to different
regimes of coexistence, as determined by the number of maxima of
$P(N_A)$:\newline
\noindent\textit{Monodominance}: $P(N_A)$ is U-shaped, achieving its
two maxima at $N_A=0$ and $N_A=N$; \newline
\noindent\textit{Mixed}: $P(N_A)$ has three maxima at $N_A=0$ and
$N_A=N$ as before plus $N_A=N/2$, meaning alternation between states of
monodominance and pure coexistence as defined below;\newline
\noindent\textit{Pure coexistence}: $P(N_A)$ is bell-shaped and has a
  single maximum at $N_A=N/2$, so that most of the time the
  populations fluctuate around the equipopulated state.

\begin{figure}[t!]
\centering
\strut\hspace{-.4cm}\includegraphics[width=.48\textwidth]{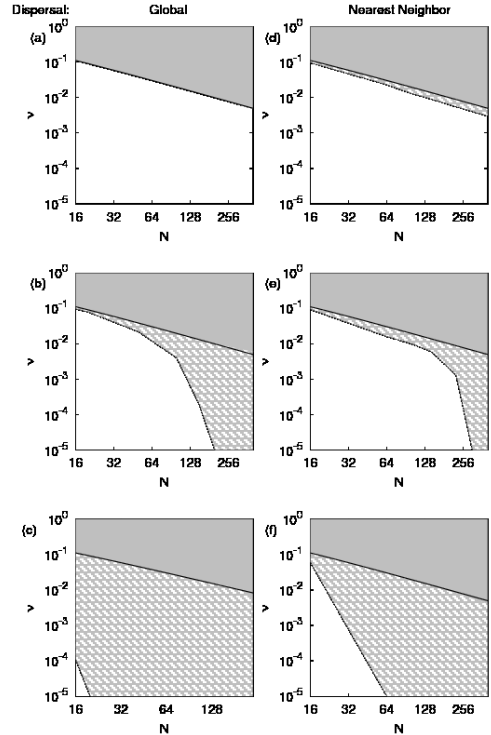}
\caption{Phases in the $N-\nu$ parameter space. White regions denote
  monodominance, gray-filled regions pure coexistence, and dashed
  regions mixed phases.  On the left (panels a-c) data obtained with
  the global dispersal model, while on the right (panels d-f) with the
  nearest-neighbor model. The three rows correspond to different
  values of $\gamma=0,\ 0.3,\ 1.0$ (from top to bottom). Axis are in
  log scale. Notice that for $\gamma=1$ and global dispersal (panel c)
  we could not show the curves for $N>240$ due to lack of statistics.}
\label{figphases}
\end{figure}

We quantify the stabilizing effect of habitat preference by studying
for which values of the parameters $\gamma$, $\nu$, and $N$, the above
defined regimes are observed.  Figure~\ref{figphases} shows, for
different values of $\gamma$, how the monodominance (white),
coexistence (gray) and mixed (patterned) regimes organize in the
$N-\nu$ plane for both global (left) and nearest-neighbor (right)
dispersal.

As observed for the extinction times, nearest-neighbor and global
dispersal models display qualitatively similar features. The only
difference is that in the neutral case $\gamma=0$ the global dispersal
model never shows a mixed distribution; conversely, with
nearest-neighbor dispersal, a tiny region of mixed phase exists also
for $\gamma=0$, compare Fig.~\ref{figphases}a and d. For $\gamma=0$
and global dispersal, the transition between monodominance and
coexistence can be analytically obtained and occurs for
$\nu>\nu_c(N)=2/(2+N)$ (see Appendix~\ref{appendix}).

\begin{figure}[b!]
\centering
\includegraphics[width=.45\textwidth]{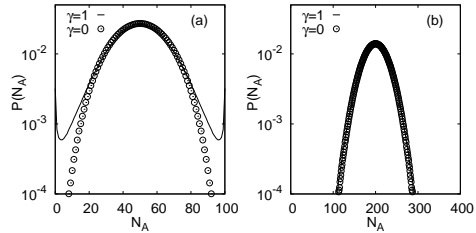}
\caption{Non-neutral vs neutral  distribution of population
$P(N_A)$: (a) for $N=100$ with $\gamma=1$, $\nu=0.005$ (solid) and
$\gamma=0$, $\nu=0.12$ (symbols): (b) for $N=400$ with
$\gamma=1$, $\nu=0.001$ (solid) and $\gamma=0$, $\nu=0.123$
(symbols). In both cases, the $y$ axis is in logarithmic
scale.\label{figcomp}}
\end{figure}

The most surprising feature is that the critical immigration rate
$\nu_c(N)$ for observing coexistence seems to be independent of
$\gamma$ and of the dispersal. Indeed, we could not determine
appreciable quantitative differences between the two extrema of global
and nearest-neighbor dispersal. In all cases, the transition to
coexistence occurs for values of the immigration rate well described
by the neutral prediction, $\nu_c(N) \approx 2/(2+N)$. Unfortunately,
we could not systematically explore larger values of $\gamma$ or $N$
as it requires huge statistics for distinguishing between the mixed
and coexistence regimes. In fact at increasing $\gamma$ and/or $N$ the
distribution becomes strongly peaked on $N/2$ with very low
probabilities on the tails, so that the region where the differences
between mixed and pure coexistence manifest is inaccessible (see
Fig.\ref{figcomp}).

Summarizing, the main effect of increasing $\gamma$ (from top to
bottom in Figure~\ref{figphases}) is to stabilize the coexistence by
increasing the portion of the $N-\nu$ plane where the mixed regime is
realized.  An equivalent analysis could in principle be performed by
plotting the transition lines in the $N-\gamma$ and/or in the
$\nu-\gamma$ plane. However, the difficulty in sampling the tails of
the distribution strongly limits the range of values of $\gamma$ and
$N$ which can be explored, so that these plots would not add much
information to Figure~\ref{figphases}).

\bigskip \centerline{\textit{Neutral vs Non-neutral: coexistence} \&
  \textit{invasibility}} \medskip We now directly compare neutral and
non-neutral dynamics by exploring whether the former ($\gamma=0$) can
reproduce/mimic the latter, e.g., in terms of generating similar
patterns of coexistence such as the distribution $P(N_A)$.  As both
$\gamma$ and $\nu$ promote coexistence, it is natural to expect that
the lack of habitat preference in the neutral model could be
compensated by a larger immigration rate $\nu$ to reproduce similar
distributions. However, as we will see, the different stabilizing
effects of immigration and habitat preference have interesting
dynamical consequences.

In particular, we consider two attempts to reproduce non-neutral
distributions with neutral ones for two different community sizes
$N=100$ (Fig.~\ref{figcomp}a) and $N=400$ (Fig.~\ref{figcomp}b). In
both cases, we proceed as follows. We fixed the habitat preference
intensity, $\gamma=1$ in this specific example. Then we simulated the
neutral version of the model ($\gamma=0$) and varied $\nu$ until the
distribution looked as similar as possible to that obtained with
habitat preference. In other words, we searched for the value of the
immigration rate $\nu$ that compensated best the absence of habitat
preference. As expected, this compensation is obtained by using a
larger value of $\nu$.

For $N=100$, the agreement between the curves is very good in the
central peak, meaning that the differences between the two models can
be appreciated only when one of the two species is dominating. For
$N=400$ the distributions are almost indistinguishable.  The fact that
the non-neutral system of Fig~\ref{figcomp}a is in the mixed phase is
clear by looking at the distribution tails; however this feature could
be difficult to detect in an experimental time series. In
Fig.~\ref{figcomp}b, the tails have such a low probability that are
essentially inaccessible even in long simulations, so that we cannot
distinguish between mixed and coexistence phases. 

For large $N$, both in the coexistence and mixed phase, the
distributions around the central mode are well fitted by a
Gaussian. In the neutral model, the limiting Gaussian distribution can
be analytically derived (see Appendix \ref{appb}), showing that the
variance in this case is proportional to $N$. Numerical simulations
(not shown) suggest that variance in the non-neutral version of the
model scales in the same way with $N$, with a prefactor rapidly
decreasing at increasing $\gamma$ reflecting the presence of niches.

Even if the difference between the models is essentially undetectable
when looking at the distributions, it can play an important role when
studying invasibility properties, i.e. by considering a situation in
which one of the species is absent at the initial time (which means to
directly test the tails of the distribution).  As an example, for the
same parameters of Fig.~\ref{figcomp} we prepared the initial state
with only species B present and we computed the average time it takes
to reach perfect coexistence ($N_A=N_B=N/2$) for the first time. When
$N=100$ (a), this time is roughly equal to $46$ generation for the
model with habitat preference and $17$ generation for the neutral one,
while for $N=400$ (b) the difference is even more pronounced ($61$ vs
$21$ generations, respectively).

\bigskip
\centerline{\textit{ Habitat preference statistics}}
\medskip

We now investigate the role of preference sites in achieving
coexistence and determining the spatial distribution of species. The
natural quantity to look at is the average fraction of individuals of
both species occupying their preference sites,
$P_{occ}=(N_{Aa}+N_{Bb})/N$. This quantity is a suitable measure of
the non-neutrality of the system: it is equal to $1/2$ in the neutral
case, as individuals have no reason to prefer one of the two site
types. The average occupation increases with $\gamma$ eventually
reaching $1$ when the distribution of sites strongly determines the
spatial distribution of the two species.

\begin{figure}[t!]
\centering
\includegraphics[width=0.45\textwidth]{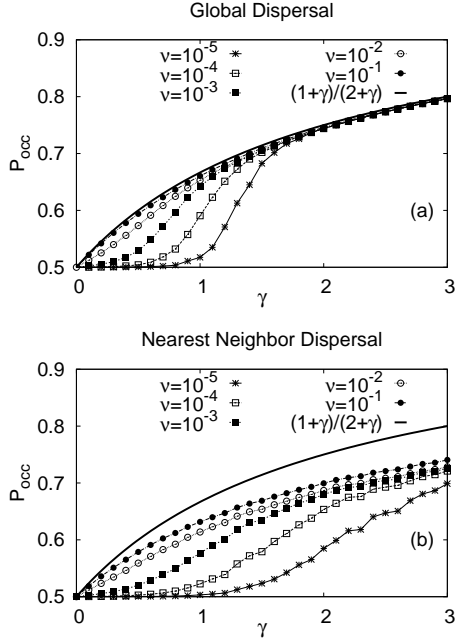}
\caption{Average preference occupation $P_{occ}$ as a function of
  $\gamma$ for $N=100$ and several values of $\nu$ as in labels. (a)
  Results obtained with global dispersal, (b)
  nearest-neighbor dispersal. Data have been averages over $10^6$
  generations. Data obtained with nearest-neighbor dispersal are also
  averaged over $200$ independent realization of the initial
  distribution of preference sites.  The continuous curve represents
  the simple prediction $P_{occ}=(1+\gamma)/(2+\gamma)$ (see
  text).}\label{figocc}
\end{figure}

The average occupation $P_{occ}$ as a function of $\gamma$ is shown in
Fig.~\ref{figocc} .  For illustrative purposes, here, we fixed $N=100$
and varied $\nu$. As one can see the curves are weakly dependent on
the choice of $\nu$, especially when $\gamma$ is large.  The
dependence on $N$ is not shown but it is also rather weak. We remark
that in the nearest neighbor case it is crucial to average over
different realizations of the dynamics and of the distribution of
advantageous sites: different spatial arrangements of the sites may be
harder (or easier) to occupy for one of the species.

When the two species are exactly equipopulated and with global
dispersal, one can easily derive from (\ref{eq:globaldisp}) that the
average occupation must be equal to $(1+\gamma)/(2+\gamma)$. This
prediction is shown for comparison in Fig.~\ref{figocc}a.  Significant
deviation are present for small $\gamma$ and $\nu$ values, for which
pure coexistence is not achieved and species are not equipopulated,
i.e. the system is in the monodominance or mixed phase.  However, it
must be noticed that with nearest-neighbor dispersal
(Fig.~\ref{figocc}b) even for large $\gamma$ the prediction
$(1+\gamma)/(2+\gamma)$ is never realized due to dispersal limitation,
which can prevent species from colonizing advantageous sites. In this
case the average occupation does not reach $1$ even for very large
$\gamma$. The presence of patchiness in the distribution of preference
sites is expected to enhance this effect: one could encounter
situations in which a whole patch is unreachable because it is
surrounded by patches being advantageous to the other species. On the
other hand, a longer dispersal range could compensate for the presence
of patches. In other words, the possibility for the occupation to
reach one for large $\gamma$ will depend on the ability of species to
reach all the favorable patches, which is known to be a crucial
feature in fragmented landscapes \citep{Keitt1997}.

We conclude this section by discussing whether the results of
Fig.~\ref{figocc} are determined by the particular choice of the
ecological advantage. To check the robustness of our results, we
simulated the variant of the model in which the advantage gives rise
to a lower mortality in the preference sites, instead of a larger
colonization probability, obtaining very similar curves to those of
Fig~\ref{figocc}.

\section{Discussion}

We have introduced an individual-based model of competition between
two species in the presence of habitat preference. As is typical in
stochastic models of community assembly, the system drifts towards the
extinction of one of the two species when immigration from the
metacommunity is not taken into account
\citep{Macarthur1967,Loreau1999}. Nevertheless, even without
immigration, habitat preference has a strong stabilizing effect on the
coexistence in large communities, leading to a dramatic increase of
the average time for the extinction of a species, consistently with
studies of niche models \citep{Tilman2004,Adler2007,Lin2009}. In
particular, \citet{Lin2009}, extending previous results
\citep{Zhang1997,Yu1998,Wright2002}, introduced a model where both
mortality and fecundity were varied among the species in such a way
that the averaged life-history fitness of several species is the same,
so to fulfill the weaker requirement of the equivalence of average
fitness of the neutral theory \citep{Hubbell2006}. The presence of
birth/fecundity tradeoffs (globally equalizing the performances of
different species) has a stabilizing effect, without affecting the
ability of predicting biodiversity patterns.

The main message of the present study is that neutral and non-neutral
coexistence may yield almost identical patterns such as the population
distribution (see, e.g., Fig.\ref{figcomp}b) while dynamic properties
may be significantly different, as revealed by invasibility
experiments \citep{Daleo2009}. In this perspective, the
advantage of the proposed model is to allow for quantitatively
comparing niche and neutral competition in a simple setting and thus
for providing an insight on the origin of such differences.  It is
thus worth discussing the predictions of the model for invasibility
experiments.

The simplest setting is that of an isolated ecosystem (i.e. without
migratory flux), in which a new species with a small population is
introduced.  This corresponds to Fig.\ref{figtempi}b, where the
probability of the invading species to take over grows with $\gamma$.
The ecological interpretation is that niche-based assemblies are
easier to invade than neutral ones if the niches are not saturated,
simply because the availability of a free niche facilitates the
invasion (see \citet{Melbourne2007} for a recent review on the role of
heterogeneity on invasibility properties).  

Another possibility is to consider the invasion of an ecosystem
initially populated by a dominant species in the presence of a regular
flux of immigrants species from the metacommunity.  In this case, if
we compare neutral and non-neutral dynamics at equal immigration rate
there is hardly any difference with respect to what is observed in the
absence of immigration.

Conversely, the response may be quite different if the comparison is
made at equal realized distributions (see Fig.\ref{figcomp}).  The
neutral model realizes patterns of coexistence similar to non-neutral
ones by compensating with a larger immigration pressure. Under these
circumstances, if some catastrophic event leads to the extinction of
one of the species, the non-neutral dynamics takes longer than the
neutral one to recover the equipopulated state. Therefore, non-neutral
coexistence is more robust, but it is also harder to achieve. The
presence of two habitats makes both the two species coexist and one
species hard to invade; in other words, niche-based coexistence is
history-dependent. An important practical consequence is that
inferring immigration rates from the fit of the distributions with
neutral models could result in overestimates, even in situations in
which the quality of the fit is very good.

The above result agrees with the observation that large scale,
species-rich patches are easier to invade
\citep{Robinson1995,Stohlgren2003}, contrarily to the prediction of
theories of competitive exclusion via niche partitioning and data from
small scale observations \citep{Tilman1997}. These contrasting
evidences constitute the so-called ``invasion paradox''
\citep{Fridley2007}. Neutral theory predicts that the chances of
establishment of an alien species depend on the frequency of
introduction of new individuals only \citep{Daleo2009}. On the other
hand, in a non-neutral community this chance is strongly influenced by
habitat diversity and availability of free niches. Our results suggest
that, when comparing the invasibility properties of ecosystems, it is
crucial to establish in which measure the diversity is maintained by
the intensity of the immigration flux and/or by niche availability.
The two-species framework considered here the stationary solutions
have not the same complexity as those of multi-species
models. However, previous studies
\citep{Chave2002,McGill2003,Tilman2004} also pointed out that species
abundance distributions of neutral and niche-based multi-species
models are very similar, provided suitable parameters are tuned and
suggesting that at least the qualitative features of invasibility
dynamics here identified should hold also in the multi-species case.
We recall that in multi-species models speciation, as well as
immigration, contributes to the introduction of new species,
especially when large systems are considered. Therefore, the problem
of inferring these rates from observed patterns becomes even more
dramatic, since it is hard to compare these estimate with reliable
measures of speciation rates. This could explain the high speciation
rates predicted by neutral theory (see \citet{Ricklefs2003} and also
\citet{Hubbell2003,Etienne2009}).

Finally, the occupation probability provides a measure of the
departure from the static equilibrium case in which each species is
confined to its preference sites. It is known that models correlating
species distributions with environmental and climatic features cannot
fully determine the observed geographical range of species
\citep{Svenning2004}.  In fact, a complete predictive power of
species-distribution models would require the existence of a static
equilibrium state \citep{Guisan2005}, an assumption that is directly
challenged by the neutral theory and partially violated by
niche-neutral models like the one presented in this paper. The
departure from the equilibrium case is enhanced by dispersal
limitation, as shown by the comparison between the nearest neighbor
case and the global dispersal of Fig. \ref{figocc}. Indeed, this
observable is the one showing the most significant dependence on the
choice of the dispersal among those considered in this work.  This
difference will become more dramatic if one adopts the more realistic
choice of a correlated environment, i.e. a distribution of preference
sites being correlated in space.  As it has been studied for the case
of fragmented landscapes \citep{Keitt1997}, the comparison of
dispersal ranges and characteristic size of patches determines whether
species will be able to reach all their preference sites or will be
arrested by dispersal limitation.

Summarizing, we studied the effect of habitat preference on the
stochastic competition between two species.  In the absence of
immigration, demographic stochasticity eventually leads to the
extinction of one of the two species. However, habitat preference
leads to a dramatic increase of the extinction time.  The drift to
extinction is arrested by introducing an immigration rate.  When the
latter is large enough, the model generates a coexistence state in
which both species occupy a significant fraction of the system.  By
compensating with a larger immigration rate the loss of specificity,
the neutral case ($\gamma=0$) can generate distributions which are
essentially indistinguishable from those obtained in the presence of
habitat preference ($\gamma>0$). In spite of the pattern similarity,
the dynamics of the two cases is very different. When only one species
is present at the beginning of the simulation, the neutral variant of
the model is much easier to invade than the non-neutral one.  This
latter result reproduces in a simplified framework what is observed in
species-rich community models: neutral and non-neutral models
reproduce similar patterns of species abundances, here of population
distribution, but can be distinguished by looking at dynamical
properties.

\section{Acknowledgments}
 We thank P. Gerlee, K. H. Andersen, M. A. Mu\~{n}oz and L. Bach for
comments on a preliminary version of the manuscript.

\appendix

\section{Condition for coexistence in the neutral model
  with global dispersal\label{appendix}}

For $\gamma=0$, i.e. in the absence of habitat preference the model
with global dispersal reduces to the Moran model with mutation
\citep{Moran1958}. In this case the transition rates can be expressed
in terms of a single population $N_A$.  In particular, equations
(\ref{eq:globaldisp}) reduce to the birth and death rates of
population $N_A$, i.e. $\mathcal{W}_{N_A\! \to N_A\pm 1} =
W^{\pm}(N_A)$ which, incorporating the effect of immigration, read
\citep{Karlin1962}:
\begin{equation}
\label{neutralrates}
\begin{array}{ll}
W^+(N_A)=&\frac{N-N_A}{N}\left[(1-\nu)\frac{N_A}{N}+
\frac{\nu}{2}\right]\\
\\
W^-(N_A)=&\frac{N_A}{N}\left[(1-\nu)\frac{N-N_A}{N}+\frac{\nu}{2}\right]
\end{array}.
\end{equation}
These rates can be used to calculate the equilibrium distribution
$P(N_A)$ through the detailed-balance relation
\begin{equation}
\frac{P(N_A+1)}{P(N_A)}=\frac{W^+(N_A)}{W^-(N_A+1)}\,,
\end{equation}
which must hold at stationarity since the process is one dimensional
\citep{Gardiner2004}.  The above expression can be used to determine
the critical value of the immigration rate for the transition from
monodominance to coexistence. It is enough to determine the value of
$\nu$ for which $P(N_A)$ passes from being a decreasing function of
$N_A$ (we are assuming $N_A<N/2$) to an increasing one.  Coexistence
is thus obtained when $P(N_A+1)>P(N_A)$, which using
(\ref{neutralrates}) and after some algebra can be recast as
\begin{equation}
[(2+N)\nu-2](N-2N_A-1)>0\,.
\end{equation}
Therefore, whenever $\nu>2/(2+N)$ the distribution is increasing up to
$N_A=(N-1)/2$ then decreasing, i.e.  there is coexistence. When
$\nu<2/(2+N)$ the opposite happens and there is monodominance.  We
conclude noticing that for $\nu=2/(2+N)$, i.e. at the the transition
between the two classes of distributions, the distribution becomes
uniform, i.e. $P(N_A)=1/N$. This is true only for the neutral model
with global dispersal.

\section{Gaussian limit of the neutral model}\label{appb}

In this Appendix we show that the stationary solution approaches a
Gaussian in the large $N$ limit, at least in the neutral, global
dispersal case. We perform the calculation in the diffusion
approximation; the procedure is similar to the classic results for the
Moran model, see, e.g. \citep{Karlin1962}, apart from small
differences in how the immigration (mutation) mechanism is defined.
From the rates (\ref{neutralrates}) the diffusion approximation yields
the Fokker-Planck equation:
\begin{equation}
\partial_t \rho(x,t)= -\partial_x (a(x)\rho(x,t))+\frac{1}{2} \partial^2_x (b^2(x)\rho(x,t)) 
\end{equation}
where $x=N_A/N$ and
\begin{equation}
a(x)= \frac{\nu(1-2x)}{2};\qquad
b^2(x)= \frac{2x(1-x)(1-\nu)+\nu}{N}.\end{equation}
The stationary distribution is easily found to be
\begin{equation}
\rho_s(x)\propto \left[2x(1-x)(1-\nu)+\nu\right]^{\frac{N\nu}{2(1-\nu)}-\frac{1}{2}}.
\end{equation}
When $N$ is large, an expansion around the maximum $x_0=1/2$ leads to the
Gaussian distribution:
\begin{equation}
\rho_{st}(x)\sim \frac{1}{\sqrt{2\pi \sigma^2}}\exp\left[-\frac{(x-x_0)^2}{2\sigma^2}\right]
\end{equation}
with $\sigma^2=(1+\nu)/[4(N+1)\nu-4]$. This shows that the relative
fluctuations around the maximum have the expected characteristic size
$1/\sqrt{N}$.


\begin{thebibliography}{}
\bibitem[Adler et al.(2007)]{Adler2007} Adler, P., J. Hillerislambers, and
  J. Levine. 2007. A niche for neutrality. Ecol. Lett. 10, 95--104.

\bibitem[Alonso et al.(2006)]{Alonso2006} Alonso, D., R. S. Etienne,
  and A. J. McKane. 2006. The merits of neutral theory. Trends
  Ecol. Evol.  21, 451--457.

\bibitem[Amarasekare(2003)]{Amarasekare2003} Amarasekare,
  P. 2003. Competitive coexistence in spatially structured
  environments: a synthesis.  Ecol. Lett. 6, 1109--1122.


\bibitem[Bolker and Pacala(1999)]{Bolker1999} Bolker B. M., and
  S. W. Pacala. 1999. Spatial Moment Equations for Plant Competition:
  Understanding Spatial Strategies and the Advantages of Short
  Dispersal. Am. Nat. 153, 575–-602.

\bibitem[Beckage and Clark(2003)]{Beckage2003} Beckage, B., and
  J. S. Clark. 2003. Seedling survival and growth of three forest tree
  species: the role of spatial heterogeneity. Ecology
  84, 1849--1861.

\bibitem[Bell(2001)]{Bell2001} Bell, G. 2001. Neutral
  Macroecology. Science 293, 2413--2418

\bibitem[Brown et al.(1995)]{Brown1995} Brown, J. H., D. W. Mehlman,
  and G. C. Stevens. 1995. Spatial Variation in Abundance. Ecology
  76, 2028--2043.

\bibitem[Chase and Leibold(2003)]{Chase2003} Chase, J. M., and
  M. A. Leibold. 2003. Ecological Niches: Linking Classical and
  Contemporary Approaches. The University of Chicago Press.

\bibitem[Chase(2005)]{Chase2005} Chase, J. M. 2005. Towards a really
  unified theory for metacommunities. Funct. Ecol. 19, 182--186.

\bibitem[Chave et al.(2002)]{Chave2002} Chave, J.,
  H. C. Muller-Landau, and S. A. Levin.  2002. Comparing Classical
  Community Models: Theoretical Consequences for Patterns of
  Diversity.  Am. Nat. 159, 1--23.

\bibitem[Chave(2004)]{Chave2004} Chave, J. 2004. Neutral theory and community
  ecology. Ecol. Lett. 7, 241--253.


\bibitem[Chesson and Warner(1981)]{Chesson1981} Chesson, P., and
  R. R. Warner. 1981. Environmental variability promotes coexistence
  in lottery competitive systems.  Am. Nat. 117, 923–943.

\bibitem[Chesson (1982)]{Chesson1982} Chesson, P. 1982. The
stabilizing effect of a random enviroment.  J. Math. Bio. 15, 1–36.


\bibitem[Chesson(2000)]{Chesson2000} Chesson P. 2000. Mechanisms of
  manteinance of species diversity. Annual Review of Ecology and
  Systematics 31:343–66.


\bibitem[Cox and Griffeath(1986)]{Cox1986} Cox, J. T, and
  D. Griffeath.  2004. Diffusive clustering in the two dimensional
  voter model. Ann. Prob. 14, 347--370.

\bibitem[Daleo et al.(2009)]{Daleo2009} Daleo, P., J. Alberti and
  O. Iribarne.  2009. Biological invasions and the neutral
  theory. Diversity and Distributions 15, 547--553.

\bibitem[Durrett and Levin(1996)]{Durrett1996} Durrett, R., and
  S. A. Levin. 1996. Spatial Models for Species-Area Curves. J.  Theor. Biol. 179, 119–-127.

\bibitem[Frachebourg et al.(1995)]{Redner1} Frachebourg, L.,
P.L. Krapivsky and S. Redner. 1995. Heterogeneous Catalysis on a
Disordered Surface. Phys. Rev. Lett. 75:2891--2894.


\bibitem[Fridley et al.(2007)]{Fridley2007} Fridley, J. D.,
  J. J. Stachowicz, S. Naeem, D. F. Sax, E. W. Seabloom, M. D. Smith,
  T. J. Stohlgren, D. Tilman, and B. Von Holle. 2007. The invasion
  paradox: reconciling pattern and process in species
  invasions. Ecology 88, 3--17.

\bibitem[Gardiner(2004)]{Gardiner2004} Gardiner, C. W. 2004. Handbook of
Stochastic Methods: for Physics, Chemistry and the Natural
Sciences. Springer, New York.

\bibitem[Gilbert and Lechowicz(2004)]{Gilbert2004} Gilbert, B., and
  M. J. Lechowicz. 2004. Neutrality, niches, and dispersal in a
  temperate forest understory. Proc. Natl. Acad.  Sci. 101,
  7651--7656.

\bibitem[Gillespie(1994)]{Gillespie1994} Gillespie, J. H. 1994. The
  causes of molecular evolution. Oxford University Press Inc., USA.

\bibitem[Gravel et al.(2006)]{Gravel2006} Gravel, D., C. D. Canham,
  M. Beaudet, C. Messier. 2006.  Reconciling niche and neutrality: the
  continuum hypothesis. Ecol. Lett. 9, 399--409.

\bibitem[Guisan and Thuiller(2005)]{Guisan2005} Guisan, A., and
  W. Thuiller. 2005. Predicting species distribution: offering more than
  simple habitat models. Ecol. Lett. 8, 993--1009

\bibitem[Haegeman and Etienne(2009)]{Etienne2009} Haegeman, B., and
  R. S. Etienne. 2009. Neutral Models with Generalised
  Speciation. Bull. Math. Bio. 71, 1507--1519.

\bibitem[Holley and Liggett(1975)]{Holley1975} Holley, R. A., and
  T. M. Liggett. 1975.  Ergodic Theorems for Weakly Interacting
  Infinite Systems and the Voter Model.  Ann. Prob.,
  3, 643--663.

\bibitem[Hubbell(1979)]{Hubbell1979} Hubbell, S. P. 1979. Tree
  dispersion, abundance and diversity in a dry tropical
  forest. Science 203, 1299--1309

\bibitem[Hubbell(2001)]{Hubbell2001} Hubbell, S. P. 2001. The Unified
  Neutral Theory of Biodiversity and Biogeografy.  Princeton
  University Press, Princeton, NJ.

\bibitem[Hubbell(2003)]{Hubbell2003} Hubbell, S. P. 2003. Modes of
  speciation and the lifespans of species under neutrality: a response
  to the comment of Robert E. Ricklefs. Oikos 100, 193--199.

 \bibitem[Hubbell(2006)]{Hubbell2006} Hubbell, S. P. 2006.  Neutral
  theory and the evolution of ecological equivalence. Ecology
  87, 1387--1398.

\bibitem[Karlin et al.(1962)]{Karlin1962} Karlin, S., 
  J. McGregor.  1962. On a genetic model of Moran.
Proc. Camb. Phil. Soc., 58, 299--311.

\bibitem[Keitt et al.(1997)]{Keitt1997} Keitt, T.H., D. L. Urban, and
  B. T. Milne. 1997. Detecting Critical Scales in Fragmented
  Landscapes. Conservation Ecology 1, 4.

\bibitem[Krapivsky(1992)]{Krapivsky1992} Krapivsky, P.L. 1992. Kinetics
  of monomer-monomer surface catalytic reactions. Phys.  Rev. A
  45, 1067-–1072.

\bibitem[Leigh et al.(2004)]{Leigh2004} Leigh, E. Jr., P. Davidar,
  C. Dick, J. Puyravaud, J. Terborgh, H. ter Steege, S. Wright.
  2004. Why do some tropical forests have so many species of trees?
  Biotropica 36, 447--473.

\bibitem[Lin et al.(2009)]{Lin2009} Lin, K., D.-Y. Zhang, and F. He.
  2009.  Demographic trade-offs in a neutral model explain
  death-rate-abundance-rank relationship. Ecology 90, 31--38.

\bibitem[Loreau and Mouquet(1999)]{Loreau1999} Loreau, M., and N. Mouquet.
  1999. Immigration and the manteinance of local species
  diversity. Am. Nat. 154, 427--440.

\bibitem[McGill(2003)]{McGill2003} McGill, BJ. 2003. Strong and weak tests of
  macroecological theory. Oikos. 102:679-–685.

\bibitem[MacArthur and Wilson(1967)]{Macarthur1967} MacArthur,
  R. H., and E. O. Wilson. 1967. The theory of island
  biogeography. Princeton University Press, Princeton, N.J.

\bibitem[Masuda et al.(2010)] {Redner2} Masuda, N., N. Gibert and
S. Redner. 2010. Heterogeneous Voter Models. preprint
arXiv:1003.0768v1 [physics.soc-ph]

\bibitem[Melbourne et al.(2007)]{Melbourne2007} Melbourne, B. A.,
H. V.  Cornell, K. F. Davies, C. J. Dugaw, S. Elmendorf, A. L.
Freestone R. J. Hall, S. Harrison, A. Hastings, M. Holland,
M. Holyoak, J. Lambrinos, K. Moore, and H. Yokomizo. 2007. Invasion in
a heterogeneous world: resistance, coexistence or hostile takeover?
Ecol. Lett. 10, 77--94.

\bibitem[Mouquet and Loreau(2003)]{Mouquet2003} Mouquet, N., and Loreau M.
2003. Community Patterns in Source‐Sink Metacommunities.
Am. Nat. 162, 544--557.

\bibitem[Moran(1958)]{Moran1958}Moran, P. A. P. 1958. Random processes
  in genetics.  Proc. Camb. Phil. Soc.  54, 60--71.

\bibitem[Pigolotti et al(2005)]{Pigolotti2005} Pigolotti, S.,
  A. Flammini, M. Marsili, and A. Maritan. 2005. Species lifetimes for
  simple models of ecologies. Proceedings of the National Academy of
  Science 102, 15747--15751.

\bibitem[Pigolotti and Cencini(2009)]{Pigolotti2009} Pigolotti, S., and
  M. Cencini. 2009. Speciation-rate dependence in species-area
  relationships. J.  Theor. Biol. 260, 83–-89.

\bibitem[Pueyo et al.(2007)]{Pueyo2007} Pueyo, S., F. He, and T. Zillio.
  2007. The maximum entropy formalism and the idiosyncratic theory of
  biodiversity. Ecol. Lett. 10, 1017–-1028.

\bibitem[Ricklefs and Schluter(1993)]{Ricklefs1993} Ricklefs,
  R. E. and Schluter, D. 1993.  Species diversity in ecological
  communities: historical and geographical perspectives. University of
  Chicago Press, Chicago, USA.

\bibitem[Ricklefs(2003)]{Ricklefs2003} Ricklefs, R. E. 2003. A comment
  on Hubbell's zero-sum ecological drift model. Oikos 100, 185--192.

\bibitem[Robinson et al.(1995)]{Robinson1995} Robinson, G. R., J.F. Quinn, 
and M. L. Stanton. 1995. Invasibility of experimental habitat in a
  California winter annual grassland.  Ecology 76, 786--794.

\bibitem[Rosindell and Cornell(2007)]{Rosindell2007} Rosindell, J.,
  and S. J.  Cornell. 2007. Species-area relationships from a
  spatially explicit neutral model in an infinite landscape. Ecol.
  Lett. 10, 586--595.

\bibitem[Stohlgren et al(2003)]{Stohlgren2003} Stohlgren, T. J.,
  D. T. Barnett, and J. T. Kartesz. 2003. The rich get richer:
  patterns of plant invasions in the United States. Front.
  Ecol. Env. 1, 11-–14.

\bibitem[Svenning and Skov(2004)]{Svenning2004} Svenning, J., and
  F. Skov. 2004. Limited filling of the potential range in European tree
  species. Ecol. Lett. 7, 565--573.


\bibitem[Tilman(1997)]{Tilman1997} Tilman, D. 1997. Community
  invasibility, recruitment limitation, and grassland
  biodiversity. Ecology 78, 81--92.

\bibitem[Tilman(2004)]{Tilman2004} Tilman, D. 2004. Niche tradeoffs,
  neutrality, and community structure: A stochastic theory of resource
  competition, invasion, and community assembly. Proc. Natl. Acad.
  Sci. 101, 10854--10861.


\bibitem[Volkov et al.(2003)]{Volkov2003} Volkov, I., J. R. Banavar,
  S. P. Hubbell, and A. Maritan. 2003. Neutral theory and relative species
  abundance in ecology. Nature 424, 1035--1037. 

\bibitem[Wright(1931)]{Wright1931} Wright, S. 1931. Evolution in
  Mendelian populations. Genetics 16, 97--159.


\bibitem[Wright(2002)]{Wright2002} Wright, S. J. 2002. Plant diversity
  in tropical forests: a review of mechanisms of species
  coexistence. Oecologia 130, 1–-14.

\bibitem[Yu et al.(1998)]{Yu1998} Yu, D.W., J. W. Terborgh, and
  M. D. Potts. 1998. Can high tree species richness be explained by
  Hubbell’s null model? Ecol. Lett.  1, 193--199.

\bibitem[Zhang and Lin(1997)]{Zhang1997} Zhang, D.-Y., and
  K. Lin. 1997. The effects of competitive asymmetry on the rate of
  competitive displacement: how robust is Hubbell’s community drift
  model? J. Theor. Biol. 188:361-–367.

\bibitem[Zillio and Condit(2007)] {Zillio2007} Zillio, T., and
  R. Condit. 2007. The impact of neutrality, niche differentiation and
  species input on diversity and abundance distributions. Oikos
  116, 931--940.




\end{thebibliography}
\end{document}